\documentclass{PoS}
\usepackage{graphics,graphicx}
\usepackage{amsmath, amssymb}
\usepackage{multirow}
\usepackage{color}
\usepackage{verbatim}
\usepackage[normalem]{ulem}
\usepackage{ulem}
\usepackage{color}
\usepackage{cancel}
\usepackage{mathtools}
\usepackage{epstopdf}

\renewcommand\sout{\bgroup \color{blue}\ULdepth=-.5ex \ULset}

\renewcommand\sout{\bgroup\color{blue} \ULdepth=-.5ex \ULset}
\def\slashchar#1{\setbox0=\hbox{$#1$}  
\dimen0=\wd0     
\setbox1=\hbox{/} \dimen1=\wd1  
\ifdim\dimen0>\dimen1   
\rlap{\hbox to \dimen0{\hfil/\hfil}} 
#1     
\else     
\rlap{\hbox to \dimen1{\hfil$#1$\hfil}} 
/      
\fi}

\newcommand{\dd}{\mathrm{d}}

\title{Chiral criticality and the role of repulsive interactions in hot hadronic matter}

\ShortTitle{Chiral criticality and the role of repulsive interactions in hot hadronic matter}

\author{\speaker{Micha\l{} Marczenko}\\
        Institute of Theoretical Physics, University of Wroc\l{}aw, PL-50204 Wroc\l{}aw, Poland\\
        E-mail: \email{michal.marczenko@uwr.edu.pl}}
\author{Krzysztof Redlich\\
        Institute of Theoretical Physics, University of Wroc\l{}aw, PL-50204 Wroc\l{}aw, Poland\\}
\author{Chihiro Sasaki\\
        Institute of Theoretical Physics, University of Wroc\l{}aw, PL-50204 Wroc\l{}aw, Poland\\}
\abstract{We explore fluctuations of the net-baryon number density in hadronic matter at vanishing chemical potential. We discuss the interplay between chiral dynamics and repulsive interactions and their influence on the properties of these fluctuations near the chiral crossover in the parity doublet model. We find that the characteristic properties of cumulants are entirely linked to the critical chiral dynamics and cannot be reproduced in phenomenological models, which account only for repulsive interactions. Consequently, a description of the higher-order cumulants of the net-baryon density requires a self-consistent treatment of the chiral in-medium effects and repulsive interactions.}

\FullConference{CPOD2021 - International Conference on Critical Point and Onset of Deconfinement\\
15-19 March 2021\\
on-line}

\begin{document}

\section{Introduction}
\label{sec:introduction}

One of the key objectives in high-energy physics is to establish the thermodynamic properties of strongly interacting matter. The ab-initio methods of lattice quantum chromodynamics (LQCD) provide reliable results on the equation of state (EoS) and fluctuations of conserved charges at the vanishing and small chemical potential
~\cite{Borsanyi:2018grb,Bazavov:2020bjn,Bollweg:2020yum}. The emergence of the quark-gluon plasma is characterized by a smooth crossover, which is linked to the chiral symmetry restoration and deconfinement of color~\cite{Aoki:2006we, HotQCD:2018pds}. At larger baryon densities the EoS of strongly interacting matter cannot be directly computed in LQCD simulations and remains an open question how to overcome the sign problem.

Fluctuations of conserved charges are known to be auspicious observable in the search for the chiral-critical behavior at the QCD phase boundary~\cite{Stephanov:1999zu}, and chemical freeze-out of produced hadrons in heavy-ion collisions~\cite{Karsch:2010ck, Braun-Munzinger:2020jbk}. In particular, fluctuations of the net-baryon number have been proposed as a probe to measure the QCD critical point in the beam energy scan (BES) programs at the Relativistic Heavy Ion Collider (RHIC) at BNL and the Super Proton Synchrotron (SPS) at CERN, as well as the remnants of the $O(4)$ criticality at vanishing and small baryon densities at the Large Hadron Collider ( LHC)  and  at RHIC ~\cite{Braun-Munzinger:2020jbk,Karsch:2019mbv}. 

The LQCD results~\cite{Aarts:2018glk} exhibit a clear manifestation of the parity doubling structure for the low-lying baryons around the chiral crossover. The masses of the positive-parity ground states are found to be rather temperature-independent, while the masses of negative-parity states drop substantially when approaching the chiral crossover temperature $T_c$. The parity doublet states become almost degenerate with a finite mass in the vicinity of the chiral crossover. Even though these LQCD results are still not obtained in the physical limit, the observed behavior of parity partners is likely an imprint of the chiral symmetry restoration in the baryonic sector of QCD. Such properties of the chiral partners can be described in the framework of the parity doublet model~\cite{Detar:1988kn, Jido:1999hd, Jido:2001nt}. The model has been applied to hot and dense hadronic matter, neutron stars, as well as the vacuum phenomenology of QCD (see, e.g.,~\cite{Dexheimer:2007tn, Zschiesche:2006zj, Benic:2015pia, Marczenko:2017huu, Marczenko:2018jui, Marczenko:2019trv, Marczenko:2020jma, Dexheimer:2012eu, Steinheimer:2011ea, Sasaki:2010bp, Yamazaki:2019tuo, Ishikawa:2018yey, Motohiro:2015taa}).


In this contribution based on~\cite{Marczenko:2020omo}, we analyze the qualitative properties and systematics of the fluctuations of the net-baryon number density and the higher-order cumulants in the parity doublet model. It is systematically examined to what extent their thermal behavior is dominated by the chiral criticality, and separately is originating from hadronic repulsive interactions.

\section{Parity doublet model}
\label{sec:pd_model}

To investigate the properties of strongly-interacting matter, we use the parity doublet model~\cite{Detar:1988kn, Jido:1999hd, Jido:2001nt}. The mean-field thermodynamic potential of the parity doublet model reads~\cite{Marczenko:2020omo}
\begin{equation}\label{eq:thermo_potential}
	\Omega = \sum_{x=\pm}\Omega_x + V_\sigma + V_\omega \textrm,
\end{equation}
where the potentials are
\begin{subequations}\label{eq:potentials_parity_doublet}
\begin{align}
	V_\sigma &= -\frac{\lambda_2}{2}\sigma^2 + \frac{\lambda_4}{4}\sigma^4 - \frac{\lambda_6}{6}\sigma^6- \epsilon\sigma \textrm,\label{eq:potentials_sigma}\\
    V_\omega &= -\frac{m_\omega^2 }{2}\omega^2\textrm.
\end{align}
\end{subequations}
where $\lambda_2 = \lambda_4f_\pi^2 - \lambda_6f_\pi^4 - m_\pi^2$, and $\epsilon = m_\pi^2 f_\pi$. $m_\pi$ and $m_\omega$ are the $\pi$ and $\omega$ meson masses, respectively, and $f_\pi$ is the pion decay constant. The kinetic part, $\Omega_x$, in Eq.~\eqref{eq:thermo_potential} reads
\begin{equation}\label{eq:kinetic_thermo}
	\Omega_x = \gamma_x \int\frac{\dd^3 p}{(2\pi)^3}\; T \left[ \ln\left(1 - f_x\right) + \ln\left(1 - \bar f_x\right) \right]\textrm,
\end{equation}
where $\gamma_\pm = 2\times 2$ denotes the spin-isospin degeneracy factor for both parity partners, and $f_x$  $(\bar f_x)$ is the particle (antiparticle) Fermi-Dirac distribution function,
\begin{equation}\label{eq:fermi_dist_nucleon}
\begin{split}
	f_x = \frac{1}{1+ e^{\beta\left(E_x - \mu^\ast\right)}} \textrm,\\
	\bar f_x = \frac{1}{1+ e^{\beta\left(E_x + \mu^\ast\right)}}\textrm, \\
\end{split}
\end{equation}
with $\beta$ being the inverse temperature, the dispersion relation $E_x = \sqrt{\boldsymbol p^2 + m_x^2}$ and the effective chemical potential $\mu^\ast = \mu_B - g_\omega \omega$. The masses of the chiral partners, $N_\pm$, are given by
\begin{equation}\label{eq:doublet_masses}
	m_\pm = \frac{1}{2} \left( \sqrt{\left(g_1+g_2\right)^2\sigma^2+4m_0^2} \mp \left(g_1 - g_2\right)\sigma \right) \textrm.
\end{equation}
From Eq.~(\ref{eq:doublet_masses}), it is clear that the chiral symmetry breaking generates only the splitting between the two masses. When the symmetry is restored, the masses become degenerate, $m_\pm(\sigma=0) = m_0$. The positive-parity state, $N_+$, corresponds to the nucleon $N(938)$. Its negative parity partner is identified with $N(1535)$. In this work, we adopt the parametrization from~\cite{Marczenko:2020omo} and analyze the contribution to thermodynamics from chiral dynamics and repulsive interactions. To this end, we analyze the fluctuations of the net-baryon number at finite temperature and vanishing chemical potential.

\section{Cumulants of the net-baryon number}
\label{sec:cumulants}

In the mean-field approximation, the net-baryon number density, as well as any other thermodynamic quantity, contains explicit dependence on the mean fields. Here, we consider only $\sigma$ and $\omega$ mean fields (cf.~Eq.~\eqref{eq:thermo_potential}), thus \mbox{$n_B = n_B\left(T, \mu_B, \sigma(T, \mu_B), \omega(T, \mu_B)\right)$}. The second-order cumulant at vanishing chemical potential has the form~\cite{Marczenko:2020omo}:
\begin{equation}\label{eq:x2_leading}
    \chi_2 = \chi_2^{\rm id}\beta_{\rm rep}\;\textrm{, with }\;\;\beta_{\rm rep} = 1 - g_\omega\frac{\partial \omega}{\partial \mu_B} \textrm,
\end{equation}
where $\chi_2^{\rm id} = \chi_2^{\rm id}\left(T, \mu_B, \sigma(T, \mu_B), \omega(T, \mu_B)\right)$ is the ideal gas expression for the net-baryon number susceptibility, and $\beta_{\rm rep}$ is the suppression factor due to repulsive interactions.
    
Depending on the details of the model, $\chi_2^{\rm id}$ in Eq.~\eqref{eq:x2_leading} contains also dependence on the $\sigma$ and $\omega$ mean fields. However, at vanishing $\mu_B$, the expectation value of $\omega$ vanishes as well, i.e., the effective chemical potential is $\mu^\ast = 0$. Thus, $\chi_2^{\rm id}$ contains only the contribution from the $\sigma$ mean field. Therefore, it encodes the information on the attractive interactions, while the information on the repulsive interactions is contained in the suppression factor $\beta_{\rm rep}$.

Similarly, one derives the higher-order cumulants and their ratios as
\begin{equation}\label{eq:xn_leading}
    \chi_{n} = \chi_{n}^{\rm id} \beta_{\rm rep}^{n-1}  + \ldots \textrm{,}\;\;\;\;\;\;\;\;\frac{\chi_n}{\chi_m} = \frac{\chi_{n}^{\rm id}}{\chi_m^{\rm id}} \beta_{\rm rep}^{n-m} \ldots\textrm,
\end{equation}
where $\chi_n^{\rm id}$ is the ideal gas expression for the n'th order cumulant. Keeping the first term provides a relatively good approximation to the full expression. The separation of the attractive and repulsive contributions persists in the approximation of the higher-order cumulants, as well as in their ratios. This allows to precisely delineate the contributions of the chiral symmetry restoration and the repulsive interaction to the critical behavior of the cumulants in the vicinity of the chiral phase transition.

\section{Results}
\label{sec:results}

\begin{figure}[t!] \centering
	\includegraphics[width=0.49\linewidth]{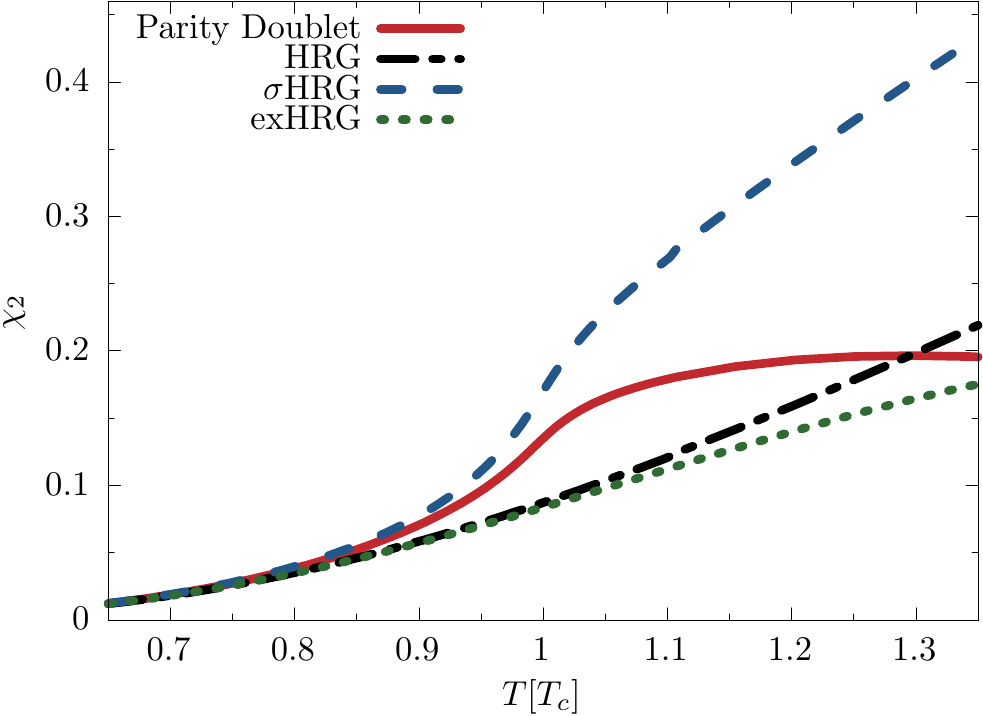}
	\caption{The second-order cumulant of the net-baryon number density from different models, see text. }
	\label{fig:cumulants}
\end{figure}

In Fig.~\ref{fig:cumulants}, we show the second-order cumulant, $\chi_2$, at vanishing baryon chemical potential. There are clear deviations of the parity doublet model result from the corresponding ideal HRG, i.e., uncorrelated gas of $N_\pm$. The influence of in-medium hadron masses is identified when considering $\chi_2^{\sigma \rm HRG}$ of the $\sigma$HRG model, where the thermodynamic potential is that of the ideal gas, but the vacuum masses of $N_\pm$ are substituted by the in-medium masses obtained by solving the parity doublet model. $\chi_2^{\sigma \rm HRG}$ increases swiftly around $T_c$, which is evidently linked to the in-medium shift of baryon masses due to chiral symmetry restoration. We note that, at vanishing chemical potential the expectation value of $\omega$ is zero, thus  $\chi_n^{\rm id}$ are equivalent to the $\sigma$HRG formulation. The result of $\sigma$HRG  model lies systematically above the ideal gas expectation. However, $\chi_2$ in HRG and $\sigma$HRG  models converge to the  Stefan-Boltzmann limit at high-temperatures. We also compare the properties of the net-baryon number cumulants with the excluded volume formulation of the repulsive interactions (labeled as exHRG). We adopt the formulation of the excluded volume effect, in which it is considered for the bulk pressure of the system~\cite{Rischke:1991ke}. For consistency, the temperature is normalized to the critical temperature, $T_c$, obtained in the parity doublet model. The exHRG results underestimate that of ideal HRG. The reduction is traced back to the repulsive interactions between hadrons. In contrast to the HRG models, the parity doublet result features a rapid increase around $T_c$, followed by a subsequent monotonic decrease to zero at high temperature. From Eq.~\eqref{eq:x2_leading}, it is clear that the difference between parity doublet model and the $\sigma$HRG an exHRG models is due to the consistent implementation of the chiral dynamics and suppression originating from $\beta_{\rm rep}$. 
	
\begin{figure}[t!] \centering
	\includegraphics[width=0.49\linewidth]{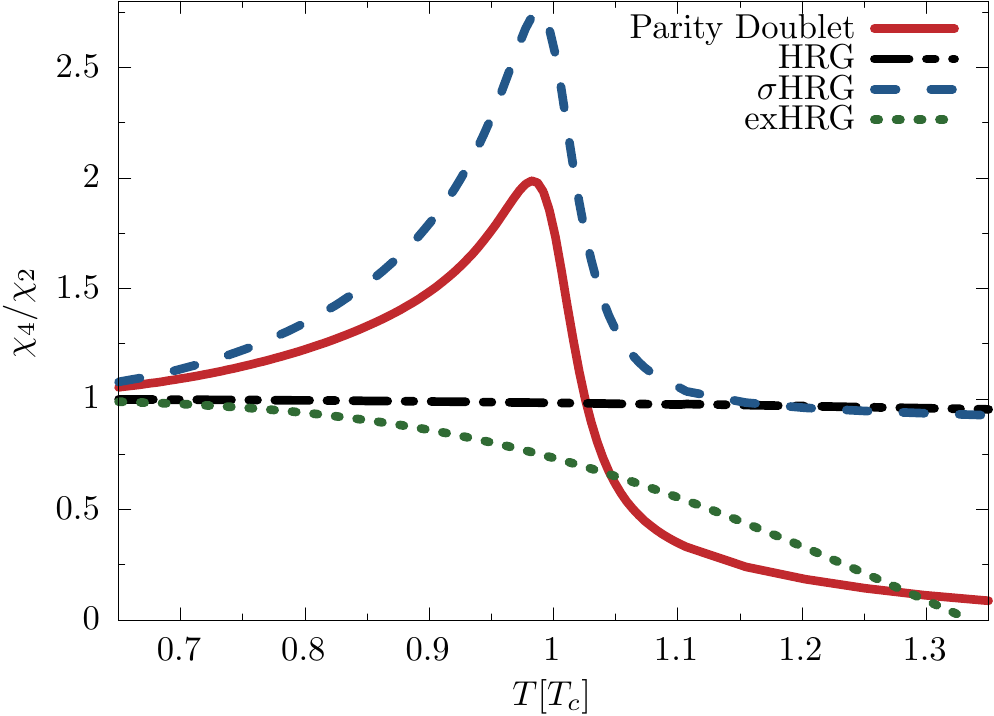}
	\includegraphics[width=0.49\linewidth]{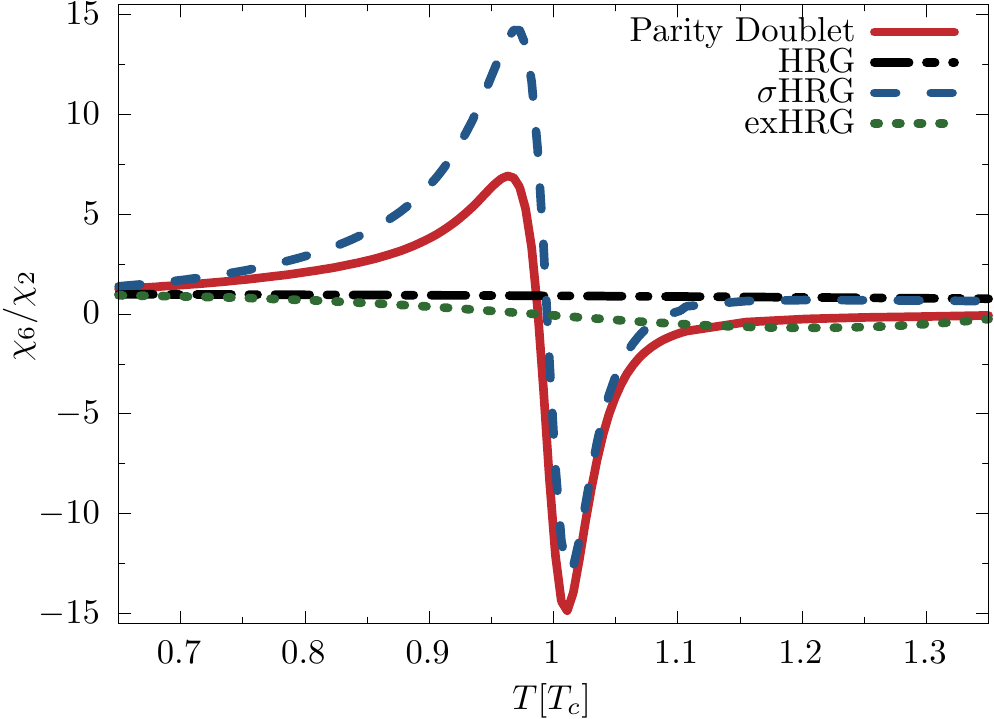}
	\caption{Ratio of higher-order cumulants of the net-baryon number density fluctuations, $\chi_4/\chi_2$ (left) and $\chi_6/\chi_2$ (right).}
	\label{fig:ratios}
\end{figure}

In Fig.~\ref{fig:ratios}, we show the net-baryon kurtosis $\chi_4/\chi_2$, and the ratio $\chi_6/\chi_2$. For the ideal HRG model, these ratios are equal to unity. The chiral dynamics and repulsive interactions implemented in the parity doublet model imply strong deviations of these fluctuation ratios from the Skellam baseline. The kurtosis exhibits a peak around the transition temperature, after which it drastically drops below unity. This is in contrast to the $\sigma$HRG result, where the peak structure appears as well, however, the result converges back to the Skellam distribution limit at higher temperatures. Thus, the appearance of the peak in the kurtosis is attributed to remnants of the chiral symmetry restoration, whereas strong suppression around $T\simeq T_c$ is due to repulsive interactions between baryons. The exHRG result shows a swift decrease from the ideal HRG behavior at low temperature and turns negative above $T_c$. 

The ratio $\chi_6/\chi_2$ exhibits a strong sensitivity to dynamical effects related to chiral symmetry restoration, as shown in the right panel of Fig.~\ref{fig:ratios}. The characteristic structure of this ratio obtained in the parity doublet model with a well-pronounced peak followed by a dip at negative values in the near vicinity of $T_c$ is expected as an imprint of the chiral criticality~\cite{Friman:2011pf}. The leading role of the chiral symmetry restoration on the properties of $\chi_6/\chi_2$ is also seen by comparing the full parity doublet and $\sigma$HRG model results in Fig.~\ref{fig:ratios}. In both cases, the structure of this ratio is preserved, albeit with some quantitative differences which are linked to repulsive interactions. Indeed, as already discussed in the context of the kurtosis, the presence of repulsive interactions suppresses $\chi_6/\chi_2$ when compared to the $\sigma$HRG results. Nevertheless, the qualitative structure of this ratio remains the same. For $\chi_6/\chi_2$ the exHRG result deviates from the ideal HRG result, turns negative, and predicts a dip above $T_c$. The parity doublet and exHRG results on kurtosis and $\chi_6/\chi_2$ are qualitatively different due to the lack of chiral in-medium effects in the latter. Therefore, such a model is not capable of reproducing the critical behavior of the fluctuations observables near the chiral crossover. We note that the exHRG behavior is also observed in the Van-der-Waals type formulation of attractive and repulsive interactions~\cite{Vovchenko:2016rkn}.

\section{Summary}
\label{sec:summary}

In this work, we systematically delineated different in-medium effects to identify the role of repulsive interactions near the chiral crossover transition within the parity doublet model in the mean-field approximation. We have carried this out by utilizing the fact that, at vanishing chemical potential, the second-order cumulant factorizes as a product of a term that is directly linked to attractive scalar interactions and a suppression factor due to repulsive interactions. Furthermore, we have found that to a good approximation, a similar separation also holds for higher-order cumulants. 
	
We have compared our results for the higher-order cumulants of net-baryon number fluctuations with an excluded volume formulation of the repulsive interactions. This model provides a substantial suppression of cumulants due to hadronic repulsion. In particular, the kurtosis  $\chi_4/\chi_2$ is reduced from unity towards the chiral crossover, as observed in LQCD results. However, when considering the $\chi_6/\chi_2$ fluctuation ratio, which exhibits a dominant contribution from the chiral criticality, such phenomenological model fails to capture the characteristic properties of this ratio. Our results indicate that in order to fully describe the properties of cumulants of net-baryon number fluctuations near the chiral crossover, it is not sufficient to account only for repulsive interactions, but it is essential to formulate a consistent framework that implements the chiral in-medium effects and repulsive interactions simultaneously.

\acknowledgments{This work was partly supported by the Polish National Science Center (NCN), under OPUS Grant No. 2018/31/B/ST2/01663 (K.R. and C.S.) and Preludium Grant No. UMO-2017/27/N/ST2\-/01973 (M.M.). K.R. also acknowledges the support of the Polish Ministry of Science and Higher Education.}


\begin{thebibliography}{}

\bibitem{Borsanyi:2018grb}
S.~Borsanyi, Z.~Fodor, J.~N.~Guenther, S.~K.~Katz, K.~K.~Szabo, A.~Pasztor, I.~Portillo and C.~Ratti,
JHEP \textbf{10}, 205 (2018)

\bibitem{Bazavov:2020bjn}
A.~Bazavov, D.~Bollweg, H.~T.~Ding, P.~Enns, J.~Goswami, P.~Hegde, O.~Kaczmarek, F.~Karsch, R.~Larsen and S.~Mukherjee, \textit{et al.}
Phys. Rev. D \textbf{101}, no.7, 074502 (2020)

\bibitem{Bollweg:2020yum}
D.~Bollweg, F.~Karsch, S.~Mukherjee and C.~Schmidt,
Nucl. Phys. A \textbf{1005}, 121835 (2021)

\bibitem{Aoki:2006we}
Y.~Aoki, G.~Endrodi, Z.~Fodor, S.~D.~Katz and K.~K.~Szabo,
Nature \textbf{443}, 675-678 (2006)

\bibitem{HotQCD:2018pds}
A.~Bazavov \textit{et al.} [HotQCD],
Phys. Lett. B \textbf{795}, 15-21 (2019)

\bibitem{Stephanov:1999zu}
M.~A.~Stephanov, K.~Rajagopal and E.~V.~Shuryak,
Phys. Rev. D \textbf{60}, 114028 (1999)

\bibitem{Karsch:2010ck}
F.~Karsch and K.~Redlich,
Phys. Lett. B \textbf{695}, 136-142 (2011)

\bibitem{Braun-Munzinger:2020jbk}
P.~Braun-Munzinger, B.~Friman, K.~Redlich, A.~Rustamov and J.~Stachel,
Nucl. Phys. A \textbf{1008}, 122141 (2021)

\bibitem{Karsch:2019mbv}
F.~Karsch,
PoS \textbf{CORFU2018}, 163 (2019)
[arXiv:1905.03936 [hep-lat]].


\bibitem{Aarts:2018glk}
G.~Aarts, C.~Allton, D.~De Boni and B.~J\"ager,
Phys. Rev. D \textbf{99}, no.7, 074503 (2019)

\bibitem{Detar:1988kn}
C.~E.~Detar and T.~Kunihiro,
Phys. Rev. D \textbf{39}, 2805 (1989)

\bibitem{Jido:1999hd}
D.~Jido, T.~Hatsuda and T.~Kunihiro,
Phys. Rev. Lett. \textbf{84}, 3252 (2000)

\bibitem{Jido:2001nt}
D.~Jido, M.~Oka and A.~Hosaka,
Prog. Theor. Phys. \textbf{106}, 873-908 (2001)






\bibitem{Dexheimer:2007tn}
V.~Dexheimer, S.~Schramm and D.~Zschiesche,
Phys. Rev. C \textbf{77}, 025803 (2008)

\bibitem{Zschiesche:2006zj}
D.~Zschiesche, L.~Tolos, J.~Schaffner-Bielich and R.~D.~Pisarski,
Phys. Rev. C \textbf{75}, 055202 (2007)

\bibitem{Benic:2015pia}
S.~Benic, I.~Mishustin and C.~Sasaki,
Phys. Rev. D \textbf{91}, no.12, 125034 (2015)

\bibitem{Marczenko:2017huu}
M.~Marczenko and C.~Sasaki,
Phys. Rev. D \textbf{97}, no.3, 036011 (2018)

\bibitem{Marczenko:2018jui}
M.~Marczenko, D.~Blaschke, K.~Redlich and C.~Sasaki,
Phys. Rev. D \textbf{98}, no.10, 103021 (2018)

\bibitem{Marczenko:2019trv}
M.~Marczenko, D.~Blaschke, K.~Redlich and C.~Sasaki,
Universe \textbf{5}, no.8, 180 (2019)


\bibitem{Marczenko:2020jma}
M.~Marczenko, D.~Blaschke, K.~Redlich and C.~Sasaki,
Astron. Astrophys. \textbf{643}, A82 (2020)

\bibitem{Dexheimer:2012eu}
V.~Dexheimer, J.~Steinheimer, R.~Negreiros and S.~Schramm,
Phys. Rev. C \textbf{87}, no.1, 015804 (2013)

\bibitem{Steinheimer:2011ea}
J.~Steinheimer, S.~Schramm and H.~Stocker,
Phys. Rev. C \textbf{84}, 045208 (2011)

\bibitem{Sasaki:2010bp}
C.~Sasaki and I.~Mishustin,
Phys. Rev. C \textbf{82}, 035204 (2010)

\bibitem{Yamazaki:2019tuo}
T.~Yamazaki and M.~Harada,
Phys. Rev. C \textbf{100}, no.2, 025205 (2019)

\bibitem{Ishikawa:2018yey}
T.~Ishikawa, K.~Nakayama and K.~Suzuki,
Phys. Rev. D \textbf{99}, no.5, 054010 (2019)

\bibitem{Motohiro:2015taa}
Y.~Motohiro, Y.~Kim and M.~Harada,
Phys. Rev. C \textbf{92}, no.2, 025201 (2015)
[erratum: Phys. Rev. C \textbf{95}, no.5, 059903 (2017)]
















\bibitem{Marczenko:2020omo}
M.~Marczenko, K.~Redlich and C.~Sasaki,
Phys. Rev. D \textbf{103}, no.5, 054035 (2021)

\bibitem{Vovchenko:2016rkn}
V.~Vovchenko, M.~I.~Gorenstein and H.~Stoecker,
Phys. Rev. Lett. \textbf{118}, no.18, 182301 (2017)

\bibitem{Rischke:1991ke}
D.~H.~Rischke, M.~I.~Gorenstein, H.~Stoecker and W.~Greiner,
Z. Phys. C \textbf{51}, 485-490 (1991)

\bibitem{Friman:2011pf}
B.~Friman, F.~Karsch, K.~Redlich and V.~Skokov,
Eur. Phys. J. C \textbf{71}, 1694 (2011)

\end{thebibliography}
\end{document}